\documentclass{article}

\usepackage{PRIMEarxiv}

\usepackage[utf8]{inputenc} % allow utf-8 input
\usepackage[T1]{fontenc}    % use 8-bit T1 fonts
\usepackage{hyperref}       % hyperlinks
\usepackage{url}            % simple URL typesetting
\usepackage{booktabs}       % professional-quality tables
\usepackage{amsfonts}       % blackboard math symbols
\usepackage{nicefrac}       % compact symbols for 1/2, etc.
\usepackage{microtype}      % microtypography
\usepackage{lipsum}
\usepackage{fancyhdr}       % header
\usepackage{graphicx}       % graphics
\graphicspath{{media/}}     % organize your images and other figures under media/ folder
\usepackage{physics}
\usepackage{algorithm}
\usepackage{algpseudocode}
\usepackage{calc}
\usepackage{indentfirst}
\usepackage{fancyhdr}
\usepackage{float}
\usepackage{mathpazo}
\usepackage{amsmath}
\usepackage{amssymb}
\usepackage{xcolor}
\usepackage{multirow}
\usepackage{array}

\title{Data-Driven Differential Evolution in Tire Industry Extrusion: Leveraging Surrogate Models
\thanks{\textit{\underline{Citation}}: 
\textbf{Garate-Perez, E., López de Calle-Etxabe, K., Ferreiro, S. Data-Driven Differential Evolution in Tire Industry Extrusion: Leveraging Surrogate Models. DOI.}} 
}

\author{
  Eider Garate-Perez \\
  Intelligent Information Systems Unit \\
  Tekniker \\
  Iñaki Goenaga 5, Eibar (20600) Spain\\
  \texttt{eider.garate@tekniker.es} \\
   \And
  Kerman López de Calle-Etxabe \\
  Intelligent Information Systems Unit \\
  Tekniker \\
  Iñaki Goenaga 5, Eibar (20600) Spain\\
  \And
  Susana Ferreiro \\
  Intelligent Information Systems Unit \\
  Tekniker \\
  Iñaki Goenaga 5, Eibar (20600) Spain\\
}

\begin{document}
\maketitle

\begin{abstract}
The optimization of industrial processes remains a critical challenge, particularly when no mathematical formulation of objective functions or constraints is available. This study addresses this issue by proposing a surrogate-based, data-driven methodology for optimizing complex real-world manufacturing systems using only historical process data. Machine learning models are employed to approximate system behavior and construct surrogate models, which are integrated into a tailored metaheuristic approach: Data-Driven Differential Evolution with Multi-Level Penalty Functions and Surrogate Models, an adapted version of Differential Evolution suited to the characteristics of the studied process. The methodology is applied to an extrusion process in the tire manufacturing industry, with the goal of optimizing initialization parameters to reduce waste and production time. Results show that the surrogate-based optimization approach outperforms historical best configurations, achieving a 65\% reduction in initialization and setup time, while also significantly minimizing material waste. These findings highlight the potential of combining data-driven modeling and metaheuristic optimization for industrial processes where explicit formulations are unavailable.
\end{abstract}

% keywords can be removed

\section{Introduction}\label{sec:intro}

Extrusion is a complex physicochemical process that is used to provide thermo-chemical properties to raw materials and is present in a wide range of industries such as polymer, food, pharmaceutical and medicine \cite{hyvarinen_modelling_2020}, \cite{zhang_3d_2021}. In a production line, the extruded materials are cooled, shaped, and cut to obtain final products \cite{rauwendaal_polymer_1986}. However, due to their complexity, extrusion processes are not easy to control and optimize \cite{sarishvili_plastic_2021}.\\

In addition to being a complex process, some industries have to manage a large variability of references. Usually, each new product change requires a stoppage of the extrusion. These interruptions together with maintenance interruptions result in a loss of optimal physical properties in the extruders, leading to a waste of material and resources. A quality decrease in the extrusion, being the first step of the production line, has a direct impact on the final product's quality. Consequently, an optimal configuration on the extruders is essential to get good quality products \cite{mount_extrusion_2017}.\\

In general, there are two trends in the field of extrusion process optimization: the works aiming at extruder design optimization and the works aiming at process optimization \cite{ma15010384}. This work focuses in the later scenario.\\

In this extrusion optimization field, the most recent research line, related to ML-based approaches, is represented by Sarishvili \cite{sarishvili_plastic_2021}. In their work, they create a digital twin (an autoencoder model) that is trained on a set of simulations created by a physical model. In their approach, they want to validate the fact that process parameters (screw speed, flow rate, and temperature in $7$ different barrels) and the output (mean residence time, torque, power engine, and specific mechanical energy) can be mapped following a hybrid modeling approach (physical model for creation observations and data-based model for learning them). Later, in \cite{burr_plastic_2023} they aim to invert the autoencoder and go one step beyond by suggesting the process input parameters that are best suited for reducing energy consumption while keeping fair product properties. Another example of process parameter analysis is the one presented by \cite{GarciaSanchez2019} where the inner and outer diameters of extruded tubes are inferred by using temperatures, pressures, and speeds of the extruder by using different data-driven models such as the k-Nearest Neighbours and Support Vector Regressors which are trained on an industrial dataset.\\

In the sight of the trend polymer extrusion research is taking, the potential of data-driven optimization using AI as a convenient approach for improving polymer extrusion processes is clear \cite{ma15010384}. Yet, most of the aforementioned works show little evidence beyond laboratory setups, note they discuss experiments \cite{sarishvili_plastic_2021}, \cite{burr_plastic_2023}, \cite{ZhangHuang2019}. Only having industrial data represented by \cite{GarciaSanchez2019}, where the existence of process parameter and output quality are related, but not optimized.
\\

At the same time, with the irruption of IoT technologies process data acquisition with sensors has become a standard in the industry. By recording the process-related information over a period of time, Artificial Intelligence (AI), and specifically Machine Learning (ML), can be used to construct models that map the process characteristics with the products' quality \cite{mai_machine_2021}, \cite{kang_machine_2020}. 
\\

One of the major obstacles in optimizing industrial problems is the lack of mathematical objective function. In a theoretical setup, this has been recently addressed in the literature with surrogate based-optimization \cite{williams_selection_2021}, metamodel-based optimization \cite{wang_evaluation_2014}, \cite{chen_time--day_2016}, surrogate-assisted optimization \cite{bartz-beielstein_high-performance_2020}, or data-driven optimization \cite{long_data-driven_2020}, \cite{pollice_data-driven_2021}. Which consists of substituting the real process by data-driven models with high accuracy, but competitive computational costs, making the optimization problem affordable \cite{remi_bardenet_surrogating_2010}.\\

Nevertheless, the recent surrogate-based optimization-related works show limited applicability in real-world optimization problems, as they are usually validated in benchmark functions. For example, \cite{yu_data-driven_2022} proposed a solution to handle high-dimension problems but was not validated in a real-world optimization problem. \cite{yang_general_2023} and \cite{dong_kriging-assisted_2021} presented a solution for constraint handling, but they were tested in benchmark functions and simple, well-defined optimization problems. \cite{zapotecas-martinez_constraint_2020} and \cite{biswas_multi-objective_2020} proposed approaches to tackle multiobjective problems, together with constraint handling and uncertainties respectively, but both have been tested in well-defined problems that were solved previously in the literature.\\ 

Surrogate optimization research lacks works that demonstrate the complexities of optimizing real-world problems with non-linear objective functions, highly correlated decision spaces, non-convex problems, unknown or computationally inefficient objective functions with constraints, small feasible regions, uncertainty, and multi-objective problems \cite{KUMAR2020100693}, \cite{GONG2022116887}.

The objective of this work is to optimize process parameters (speed settings) for a quintuplex extruder in the tire manufacturing industry using historical data. The goal is to reduce waste or rework during extrusion initialization. This study addresses physical constraints, human intervention, process usability, and production variability, which are often overlooked in extrusion optimization \cite{glinz_non-functional_2007}, \cite{OSABA2021100888} by paying special attention to the mathematical formulation of industrial scenarios. By combining data-driven models and metaheuristic algorithms under the surrogate optimization paradigm, this work aims to provide a practical solution to this complex industrial problem.\\

The paper is organized as follows. Section~\ref{sec:industry} explains the tire industry studied. Then, in Section~\ref{sec:surrogate} the methodological approach is described. Firstly, mathematical problem formulation by the use of historical data and surrogate models is described. Secondly, the proposed Data-Driven Differential Evolution with Multi-Level Penalty Functions and Surrogate Models is defined. In Section~\ref{sec:results} the results of the surrogate models and optimization algorithm are given, and these are discussed in Section~\ref{sec:discussion}. Finally, the proposal is concluded in the Section~\ref{sec:conclusions}.

\section{Industrial scenario: Rubber Extrusion Process} \label{sec:industry}

\begin{figure}[H]
    \centering
    \includegraphics[scale = 0.45]{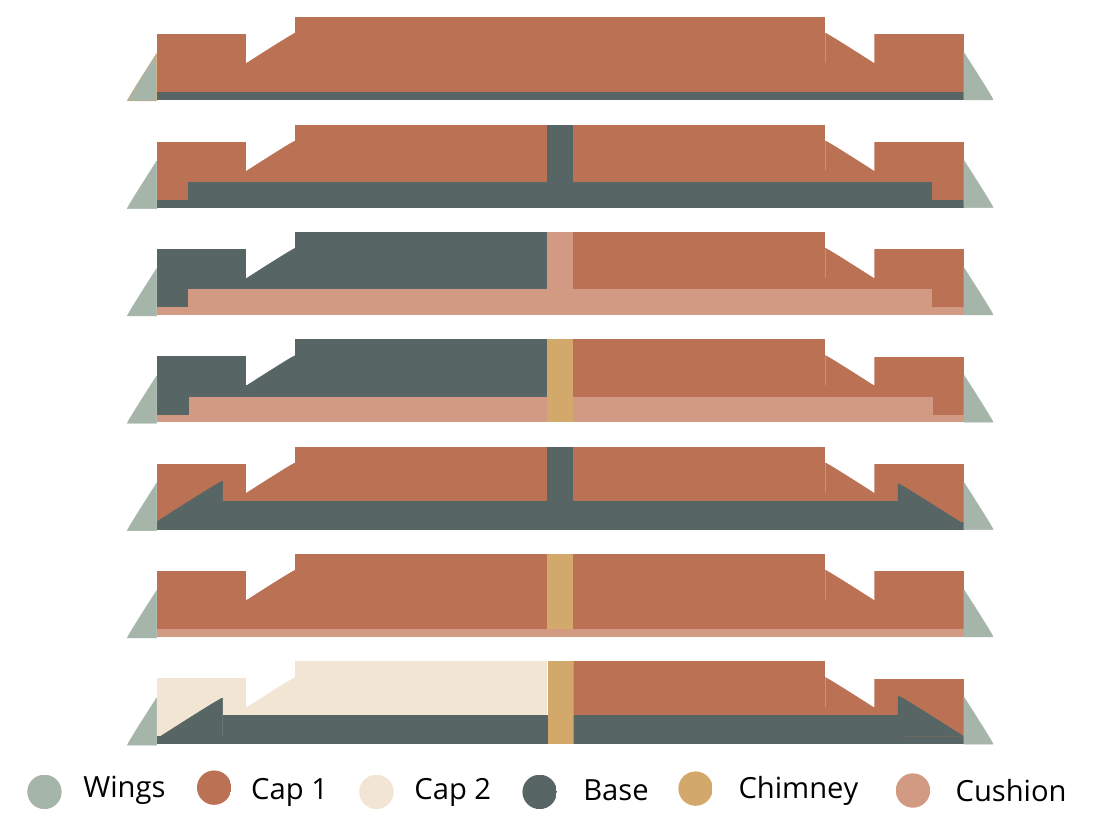}
    \caption{Different tire references created by combining various extrudates from a single quintuplex extruder. Image is inspired on the commercial flyer of Plastrading Slr.}
    \label{fig:tire_elements}
\end{figure}

    This work studies an industrial rubber extrusion process occurring in the tire manufacturing industry. This process is in charge of taking raw materials or compounds and shaping them with different dies to form extrudate products.  The process under study involves a quintuplex extruder which is an interesting type of extruder for tire industry, as by combining different extruders, the different parts of a tire can be produced (see example in Figure~\ref{fig:tire_elements}).\\

\begin{figure}[H]
    \begin{center}
        \includegraphics[scale = 0.45]{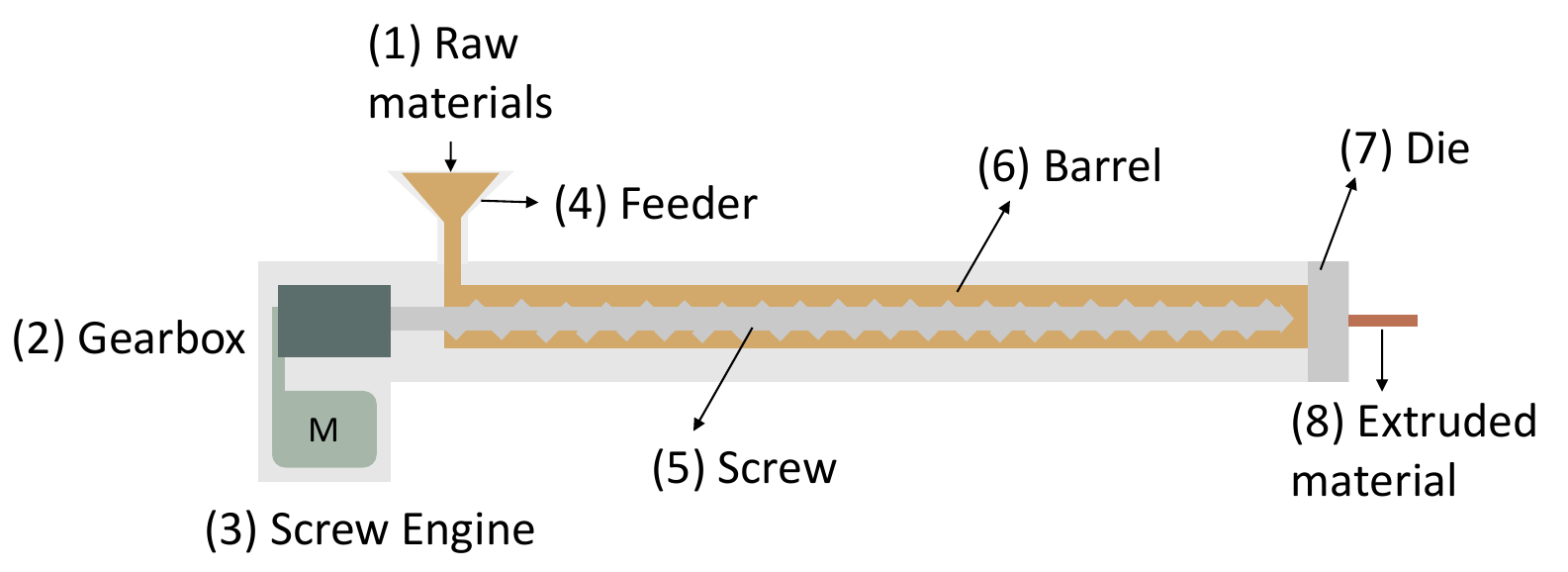}
        \caption{A general schema of an extruder with main components.}
        \label{fig:extr_schema}
    \end{center}
\end{figure}

In general, extruders present the different parts shown in Figure~\ref{fig:extr_schema}. There is a feeder (4), which is the entry point of the raw materials (1), and a screw (5), which works with an engine (3) and is used to push the material through a barrel (6) into the die (7), which is the part that shapes the material before it is released (8). Within some desirable physical conditions such as temperature, pressure, or viscosity, raw materials are transformed to come out with some target properties \cite{abeykoon_energy_2021}. 

\begin{figure}[H]
    \centering
    \includegraphics[scale = 0.45]{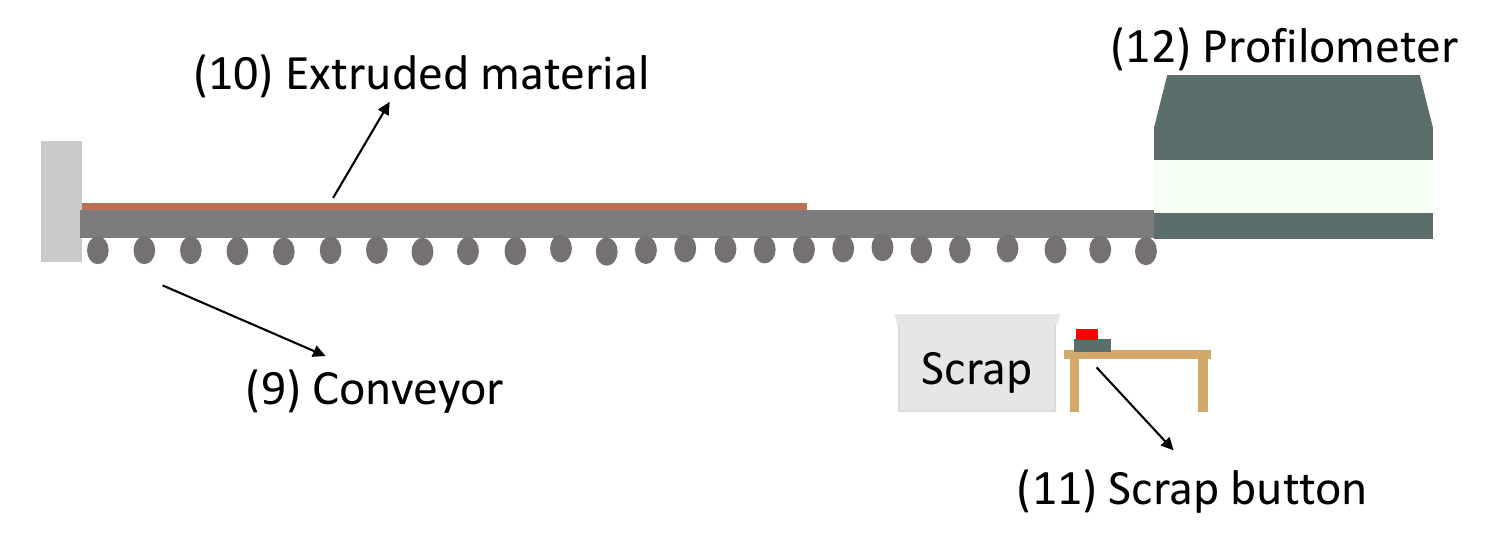}
    \caption{Visualization of the quality measurement process just after the extruded material is released.}
    \label{fig:prof_schema}
\end{figure}

In this industrial process, during each extrusion, the extruded material (10) is transported by a conveyor (9) to undergo quality testing, as shown in Figure~\ref{fig:prof_schema}. There, the quality is assessed using a profilometer (12) and the measured value is compared to a predefined quality target. This test is used as a first filter before the cooling and shaping process. In addition, there is a scrap button (11) for the human operator or automatic process control which rejects poor-quality extruded materials when activated. As in many other industrial process machines, the extruders are sensorized measuring temperatures and pressures in different points of the extruders that are used to monitor and control the extrusion process.

Besides combining the extruders and dies, operators set the extruding speeds to initialize each extrusion process. In the early stages of the extrusions, the final product exhibits low quality, requiring a brief period for the process to stabilize and produce high-quality output. This duration is related to the parameters used for initializing the extrusion and is directly correlated to the amount of scrap, the bad quality product that is produced and scrapped. In other words, the longer the time required for the process to stabilize, the greater the amount of waste created during rubber production. When evaluating product quality, three scenarios may arise: (1) the extruded material is of such poor quality that it is cut off completely before reaching the profilometer; (2) the material is not fully cut but fails to meet the required quality standards throughout the extrusion process; or (3) the material is not cut and eventually reaches the required quality after a certain period. These outcomes highlight the importance of optimal initialization, as early-stage decisions have a direct impact on material waste and product quality. 

The extrusion process is a continuous operation, where once the machinery is initialized, it remains active for an indeterminate period depending on production goals, the quality of the extruded material, and possible failures or maintenance stoppages. Consequently, in real industrial conditions, all process and quality data are collected continuously at a fixed acquisition frequency, forming time series. Importantly, each process decision—made at the moment of initialization—unfolds over time, influencing the entire subsequent evolution of the process. Therefore, for the purpose of optimization, we aim to assess the outcome of each initialization by evaluating how the process developed over time based on the decisions made. This temporal perspective is central to the problem formulation and the design of the optimization framework.

In this context, this work aims to optimize the initial machine configurations to minimize the required stabilization time and, consequently, reduce material waste during each initialization of the rubber production process. These parameters are the key variables configured at the start of the process and play a crucial role in determining the final product quality.
%%%%%%%%%%%%%%%%%%%%%%%%%%%%%%%%%%%%%%%%%%
\section{Surrogate Based Optimization}\label{sec:surrogate}

A surrogate model, or metamodel, approximates a real-life complex problem with the highest possible accuracy and with competitive and low computational cost. These models are typically employed when the original complex model is computationally expensive and/or empirical experimentation is costly \cite{CarlosCernuda1}. As outlined in Section~\ref{sec:intro}, this paper focuses on the latter scenario.

The problem is approximated by starting with a raw dataset of the studied process. This data is used to formulate the search and objective spaces of the optimization problem. In addition, three surrogate models have been employed to approximate the real extrusion process, serving as two constraints and an objective function.

Once the optimization problem is defined, it is solved using our proposed metaheuristic algorithm based on the Differential Evolutionary Algorithm \cite{noauthor_differential_2005}. This approach leverages the strengths of differential evolution to efficiently explore and exploit the search space, leading to optimal solutions for the extrusion process under the defined constraints and objective function.

\begin{figure}
    \begin{center}
           \includegraphics[scale = 0.33]{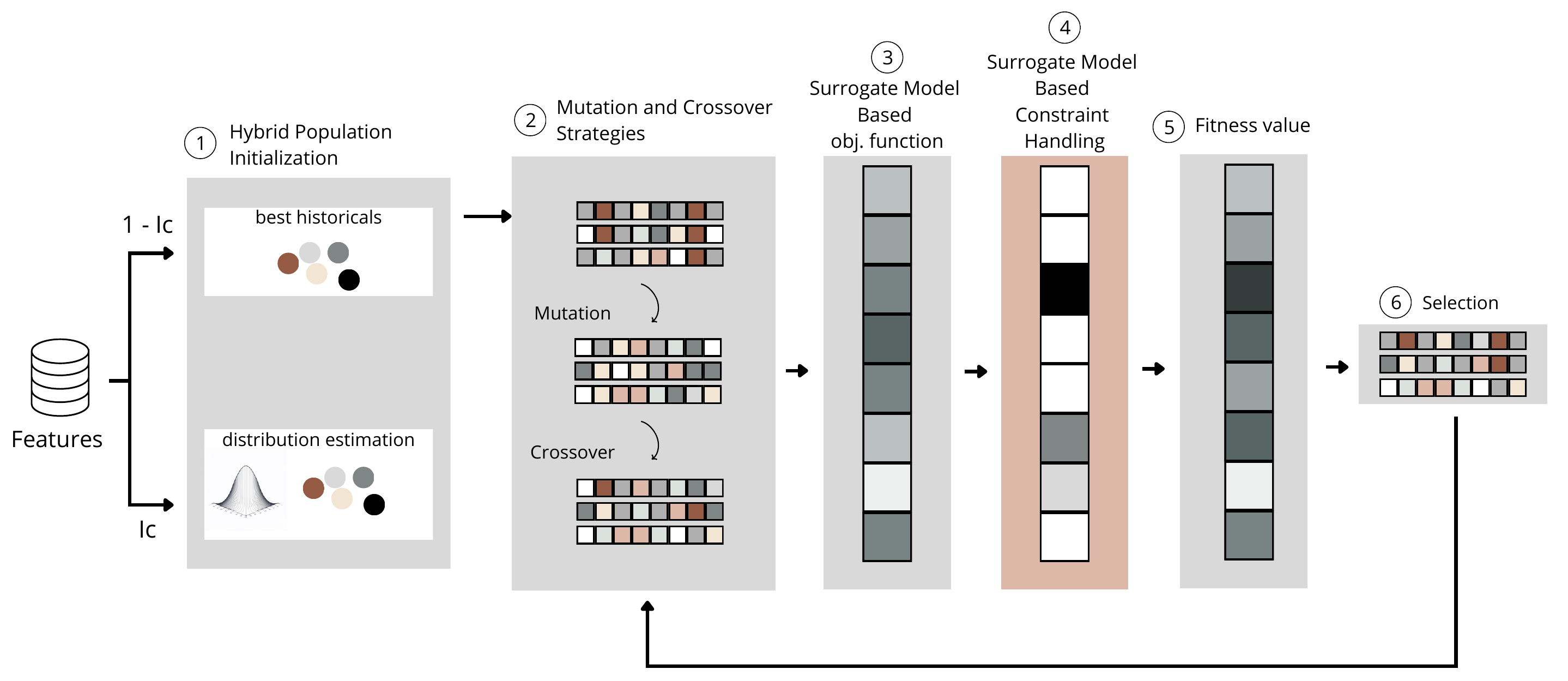}
           \caption{Graphical summary of the customized Differential Evolutionary Algorithm}
           \label{fig:MetaheuristicOverview}
    \end{center}
\end{figure}

Figure~\ref{fig:MetaheuristicOverview} presents a comprehensive diagram of our proposed optimization algorithm for extrusion process optimization. The diagram illustrates the various components of the algorithm, which are explained in this section. This visual representation highlights the integration of data-driven surrogate models and metaheuristic strategies to effectively tackle complex industrial problems.

\subsection{Surrogate-model based optimization formulation}

When optimizing an industrial problem it is essential to understand all of its details to model it properly. This requires translating all the technical aspects of the problem to be solved into a formal mathematical optimization problem.  
This key elements of an optimization problem include: the objective and decision variables, that is, the parameters that can be modified or that affect our optimization target; the objective function, the mathematical expression to minimize that establishes the relationship between the decision variables and the objective variable; and, the problem constraints, the physical limitations that bound the values the parameters can physically take. \\

In particular, this work presents a surrogate-model based optimization approach to find suitable and optimal parameters for this industrial extrusion process. In this regard, the main advantage of a surrogate model is that it enables iterating endlessly in the search for the best parameters without interfering with the complex production process. Without the surrogate model, these iterations would not be possible for the economic and temporal costs of testing with the actual production plant. At the same time, the drawback of this approach is that it relies on real industrial data that needs to be modeled and is not exact as a formal mathematical optimization problem.  

The modeling or formalization of the problem not only makes the problem solvable but also the solution depends entirely on the modeling performed. Thus, the objective is not focused on finding an optimization algorithm that discovers the best possible solution to the problem, but rather on ensuring that this best solution meets all the process requirements so that it can be implemented in the real context and is able to improve the production process effectively.

\begin{figure}[h]
    \centering
    \includegraphics[scale = 0.3]{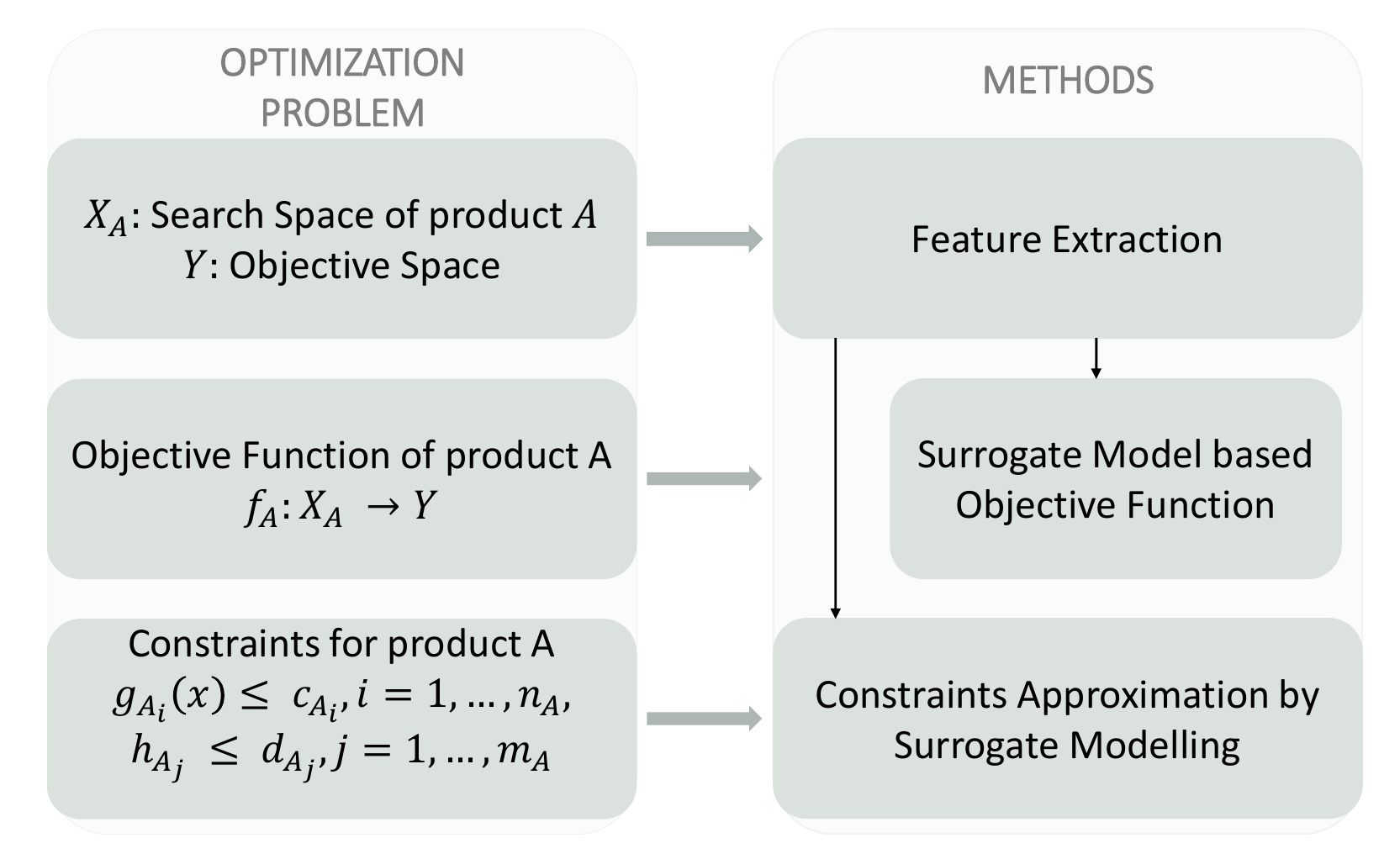}
    \caption{Schema of the approach followed in this work for the mathematical formulation of the optimization problem.}
    \label{fig:met_schema}
\end{figure}

The approach followed in this work is presented in Figure \ref{fig:met_schema} and is described in the rest of this subsection.

\subsubsection{Raw data curation}

Given that industrial data usually comes from connected machinery, pre-processing this data source is the foundation for achieving modeling of the optimization problem, as it will facilitate the definition of the search and objective spaces, as long as the training of objective and constraint functions. 

In the studied scenario the sensor readings included continuous and categorical variables. The categorical variables were used to describe the material and shapes used and were recorded every time a change was detected. And regarding the continuous data, some signals were recorded with a constant acquisition frequency while others were measured only when a magnitude change was detected with a consequent varying acquisition frequency.

To fix the acquisition inconsistencies the original raw data were curated to establish a common frequency of 1Hz, the highest frequency in the recorded signals. For that purpose the DQTS R package \cite{9845398} was used, missing values were imputed by the Last Observation Carried Forward method (LOCF), due to the low variability of the variables with lower frequencies.

This restructuring enables the measurement of properties in a time series format for each extrusion performed in the plant, paving the way for the analysis and feature engineering necessary for the definition of the problem.

\subsubsection{Search and Objective Space by Feature Extraction}

\begin{figure}[h]
    \centering
    \includegraphics[scale = 0.3]{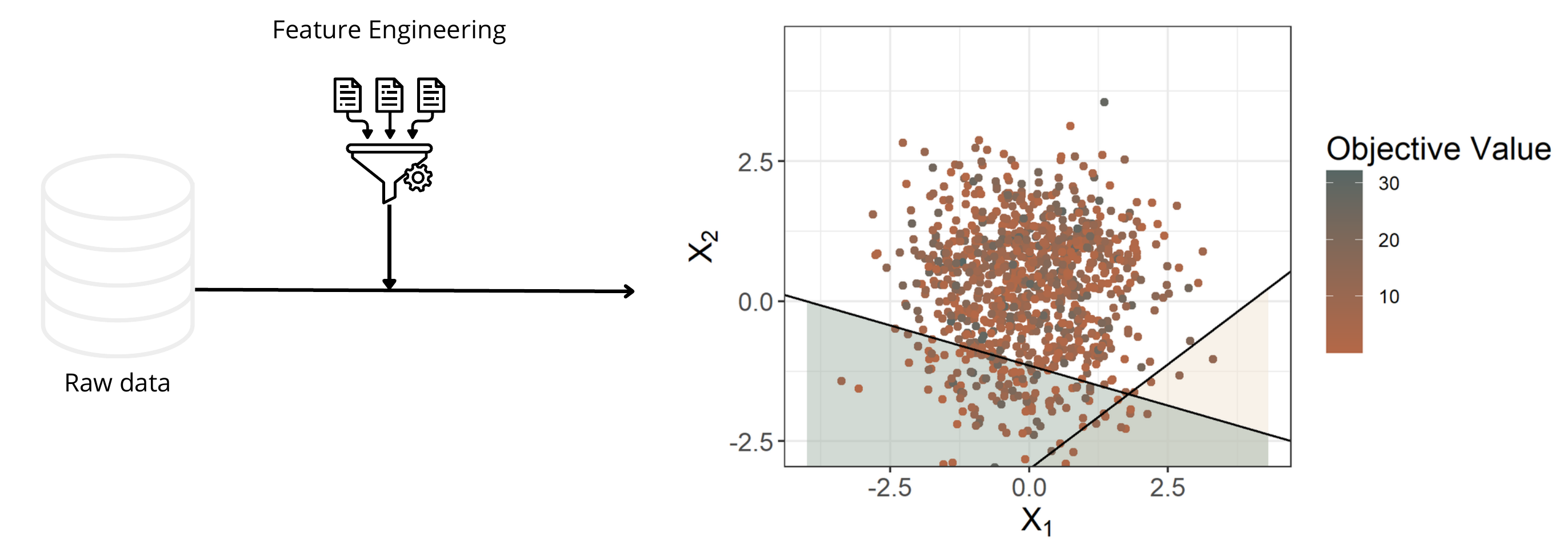}
    \caption{Seach and objective space definition by feature extraction using historical raw data.}
    \label{fig:search_space}
\end{figure}

In an optimization problem, every point of the search space (search point) needs to be represented by the same set of search variables, and, at the same time, there must exist at least one function able to quantify the goodness of these search points. In our case at hand each search point must represent an extrusion to which certain goodness is associated and must be optimized. Nevertheless, extrusions are processes of varying duration and with changing conditions. Hence, the first step is to characterize each extrusion so that the search points reflect in which conditions the extrusion was launched; and identify the goodness related to these search points. 
This characterization requires a signals segmentation and a feature extraction.

This characterization or formalization results in a dataset with sub-optimal historical solutions. This dataset is the one used in the initialization of the optimization algorithm, as seen in Figure~\ref{fig:MetaheuristicOverview} in part (1) of the scheme.\\

\begin{figure}
    \centering
    \includegraphics[scale = 0.5]{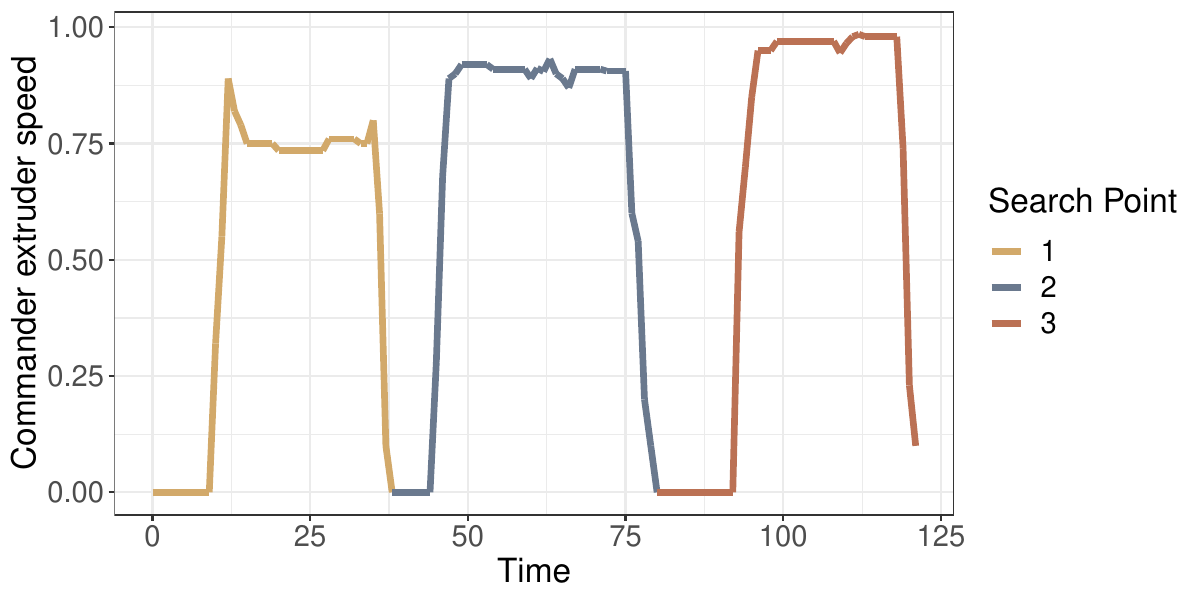}
    \caption{Segmentation of data based on commander extruder's speed. Speed values have been scaled using max-min scaler.}
    \label{fig:segmentation}
\end{figure}

\textbf{Signal segmentation}\\
To detect each extrusion process, the speed of the commander extruder was used. An extrusion is initialized when the commander extruder’s speed increases from zero, and it is finished when it drops back to zero. Given that the initialization settings can only be manipulated before initializing the extrusion each extrusion execution was defined as the stoppage time since the previous execution plus the time in active extrusion defined by the commander extruder. An example of the segmentation is depicted in Figure~\ref{fig:segmentation}.\\ 

\textbf{Feature extraction: Search space}

Once data were segmented a set of indicators (or features) was extracted from all the sensors that were available on the plant in order to define each search point value. Search variables must: be controllable by the operators; and, affect the objective variable or variables of the problem.

In the extrusion the temperatures, pressures, speeds, materials and dies were used.  A boolean was also considered to differentiate between stoppage and usage. The set of features and sensors extracted is summarized in Table~\ref{table:tabla_signals}. 

As explained above, the extracted indicators taken during the time interval prior to extrusion initialization—referred to as the stoppage period—define the search space for each data point. Specifically, the search variables consist of the minimum, maximum, and mean values of pressure and temperature, along with the target screw speed and the mean acceleration of the screw at startup.\\%honen zergatia definitu berriro?\\

\begin{table}
 \centering
 \begin{tabular}{| m{2.5cm} | m{2cm}| m{5cm} |} 

     \hline\
 \footnotesize{\textbf{Measured property}} & \footnotesize{\textbf{Position}} & \footnotesize{\textbf{Indicators}}\\
 \hline
 \footnotesize{Pressure} & \footnotesize{(4), (5)} & \footnotesize{min, max, standard deviation, mean, first and last}\\
 \hline
 \footnotesize{Temperature} & \footnotesize{(4), (5), (6)}  & \footnotesize{min, max, standard deviation, mean, first and last}\\
 \hline
 \footnotesize{Screw's Speed} & \footnotesize{(5)} & \footnotesize{Mean acceleration at the start (speed change/time nedeed), target speed}\\
 \hline
 \footnotesize{Material} & \footnotesize{(4)} & \footnotesize{First, last and change}\\
 \hline
 \footnotesize{Die} & \footnotesize{(7)} & \footnotesize{First, last and change}\\
 \hline
 \footnotesize{Usage} & \footnotesize{(5)} & \footnotesize{Usage boolean}\\
\hline
\end{tabular}
\caption{\label{table:tabla_signals} Summary of the measured properties, locations with the corresponding numbers in Figure~\ref{fig:extr_schema} and Figure~\ref{fig:prof_schema}, and used indicators in each case.}
\end{table}

\textbf{Feature extraction: Objective space}

When working on problems without prior formalizations or modeling as optimization problems, defining one or more objectives that meet process, and material standards can be complex. Not only they must meet all the aforementioned conditions and the production goals, but these variables must also be measured every time a new process or trial is conducted. Furthermore, these values need to be robust across all types of processes and products being analyzed.

In the studied process, the problems's objective was clear: to minimize the time and material waste associated with each extruder restart. Each time an extrusion is performed, it usually takes some time to reach the established quality values. The extrudate produced during this time interval does not meet the products' quality standards and is therefore usually discarded. Consequently, the restart time is directly related to the amount of material waste, implying that by minimizing the restart time, the amount of material waste is also minimized. Therefore, the studied problem is a single objective optimization problem.

Taking that premise, a variable is computed that is able to quantitatively reflect the goodness of an extrusion in terms of restart time before reaching the quality standards. However, as quality values in the process are continuously calculated over time, it is not enough to reach the predetermined value at a specific moment; the product quality must remain within those limits to prevent further rejection.

This work proposes using the time needed by the produced tread to reach the target profilometer specifications and remain steady, from now on, \textit{steadiness time}. The objective variable is computed by:

\begin{enumerate}
    \item Create a $1\%$ tolerance area around the profilometer setpoint.
    \item Compute a $5$ seconds sliding window with Root Mean Square Error (RMSE).
    \item Determine the maximum RMSE that keeps all points within the $1\%$ tolerance zone.
    \item Identify the first time instant where the RMSE from step $2$ is less than the RMSE threshold from step $3$.
\end{enumerate}

As the profilometer value is computed every time an extrusion is launched and not entirely cut (when the quality is too poor and it becomes necessary to discard the entire extrudate and restart the process), this objective variable measures the goodness of a given search point.

A simple yet sub-optimal approach to solving the problem is to select the best result from the existing dataset. However, this method is limited to the recorded results, ignores the objective functions and constraints of the problem, and cannot explore better alternatives. By approximating the constraints and objective function using Machine Learning models, it can be employed a metaheuristic algorithm that leverages the initial dataset to improve upon these results.

\subsubsection{Constraints Definition}

In optimization problems, solutions are categorized as feasible or infeasible. Constraints ensure that only physically feasible solutions are considered, guiding the algorithm towards finding the best solution among them.

After a detailed study of the extrusion process, two main constraints were identified: the compliance with minimum quality standards, and the physical consistency of the setpoints.

These constraints contribute to the fitness function in the optimization algorithm as depicted in part (4) and (5) of Figure~\ref{fig:MetaheuristicOverview}.\\

\textbf{Minimum quality constraint}

During initial explorations of \textit{steadiness time}, three types of extrusions were identified along the $5821$ labelled instances. Extrusions having a valid \textit{steadiness time} (Class $1$, 33.06\% of data) and extrusions having invalid times (Class $2$ and $3$). The cause of these invalid extrusions was tracked back to two root causes: Some extrusions did not reach profilometer steadiness (Class $2$, 7.38\% of data) and, in others, the operators cut the extruded material even before it reached the profilometer (Class $3$, 59.55\% of data). 

Given that the surrogate model would not be able to map the relation of these extrusions with their \textit{steadiness time}, from an optimization perspective, Class $2$ and Class $3$ are considered infeasible solutions. A solution is not valid for the studied optimization problem if it does not reach the established minimum quality thresholds, meaning that Class $2$ and Class $3$ are not only difficult to model but are intrinsically infeasible. To limit the range of solutions to those that are feasible (Class $1$), a supervised classification algorithm was trained.

This surrogate model will replace the visual constraint performed by the worker when formalizing the optimization. Therefore, the predictor variables for the problem must be defined by the search variables of the optimization problem. 

Due to class imbalance, before training the model, the class distribution was adjusted. Smote \cite{chawla_smote_2002} was used for the minority classes and then, an undersampling for the majority class was done using Tomek Links \cite{tomek1976two}. 

For the modeling, a Random Forest Classifier algorithm \cite{breiman_random_2001} was used. This selection was done because of the highly interactive variables space and its explainability. For the validation of the model, F1-score metric was used, which is one of the most used metrics in imbalanced classification problems \cite{bej_loras_2021}. 

All the modeling and processing for the training of the constraint was carried out using Scikit-learn and Imbalanced-learn Python libraries.\\

\textbf{Solution consistency constraint}

The \textit{steadiness time}, in addition to indicating the quality and material cost of an extrusion, also depends on the time required to reach the setpoints for temperature, pressure, and speed. Therefore, it is not feasible to achieve a \textit{steadiness time} significantly shorter than the time needed to reach these setpoints. Specifically, for speeds that start with an acceleration from zero, it is impossible to achieve a stable and high-quality extrusion without reaching the setpoint speeds.

To ensure that this relation was kept in the solutions suggested by the optimization, another constraint was introduced. 

Being $x_{speed, k}$ be the variable describing the target speed setpoint in the $k$-th extruder and $x_{acceleration, k}$ the mean acceleration at the start of $k$-th extruder, $k\in K \subseteq \{1, ..., 5\}$ where $K$ is the set of extruders in usage. Then, the time required to have the extruders speeds in the defined setpoint can be calculated as follows,

\begin{equation}\label{eq:t-set}
    t_{setpoint} = \max\{x_{speed, k}/x_{acceleration, k}\ |\ k \in K\}.
\end{equation}

Thus, the physical constraint is defined as follows,

\begin{equation}\label{eq:phy-ctr}
    t_{setpoint} < (1 + \delta)\hat{y},
\end{equation}

where $\hat{y}$ is the approximation of the \textit{steadiness time} using objective function's surrogate (see subsection \label{sec:obj-surrogate}). $\delta >0$ is the relative feasible discrepancy between $t_{setpoint}$ and $y$, which was set to $\delta = 0.5$.\\

\subsubsection{Surrogate model based Objective Function Definition}\label{sec:obj-surrogate}

The objective function is a real-valued, discrete, or mixed function that maps the decision space into the objective space which is wanted to minimize or maximize. 

As it happened with the restrictions, there was no way to explicitly define a function that estimates the \textit{steadiness time}. Therefore, we tackled the problem again using a supervised Machine Learning model, trained with the historical values of the decision space variables and the objective space variable. To differentiate between the different conditions across the different product types, the variable with the product's identification was also used as a predictor, so an unique model could be trained and used to approximate all product types. 

XGBoost \cite{chen_xgboost_2016} was the algorithm used to model the objective function. Other studies have reported the computational efficiency of the XGBoost algorithm as well as its ability to outperform other Machine Learning models \cite{ibrahem_ahmed_osman_extreme_2021}, \cite{ma_xgboost-based_2021}, \cite{li_gene_2019}.

\begin{figure}
    \begin{center}
           \includegraphics[scale = 0.4]{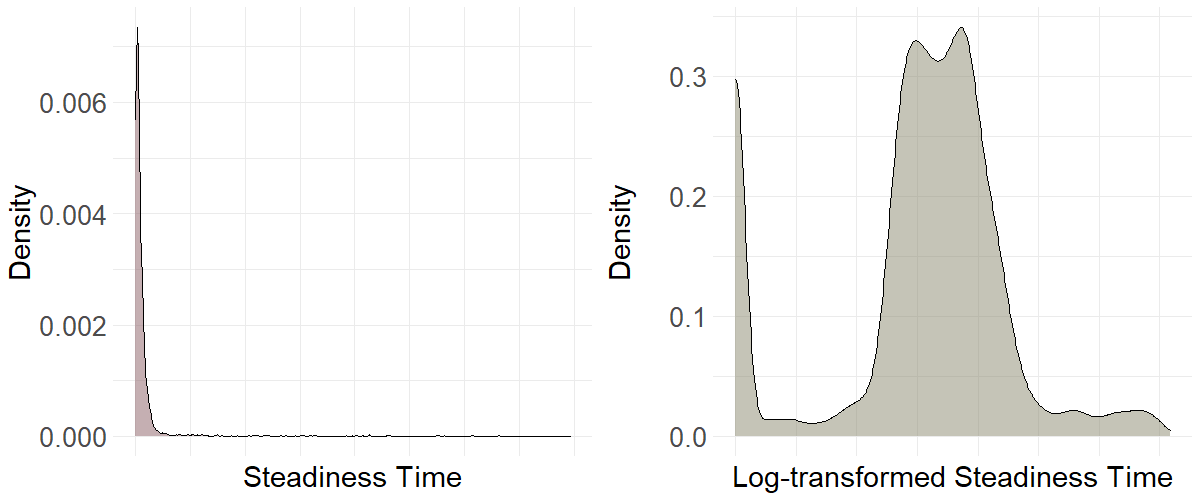} 
           \caption{\textit{Steadiness time} variable density plot and the density after logarithmic transformation.}
           \label{fig:log-normal}
    \end{center}
\end{figure}

However, XGBoost's performance can be affected by the presence of outliers \cite{dairu_machine_2021}. This can be significantly problematic if the objective variable has a log-normal distribution. In our case, \textit{steadiness time} variable, had a log-normal distribution aspect as can be appreciated in Figure~\ref{fig:log-normal}. Fortunately, log-normal distribution can be easily transformed into a normal distribution by applying the logarithmic and rejecting distribution tails as outliers. This transformation is shown in Figure~\ref{fig:log-normal}. 

The priority for the model was to be capable of ranking the extrusions' quality from good to bad, or in other words, given two possible solutions, the model had to know which of them was a better solution. Kendall Correlation Coefficient was selected for model validation as it is a statistic that measures the correlation of two ordered variables which explains the similarity of their ordering \cite{10.1093/biomet/30.1-2.81}.

The model was trained with only feasible historical observations, that is, with the extrusions that had reached the profilometer and had gotten steady within the defined range near the target value.

As this surrogate model approximates and substitutes the objective function of the problem, it took part in the optimization algorithm as it is shown in part (3) of Figure~\ref{fig:MetaheuristicOverview}.

\subsection{Data-Driven Differential Evolution with Multi-Level Penalty Functions and Surrogate Models} \label{sec:DE}

Previous sections have given the methods proposed in this study for the formalization of the problem. Once the real scenario is described as an optimization problem, optimization techniques can be used to solve them.

Metaheuristic algorithms are search or optimization algorithms that are proven to be efficient when solving complex real-world problems where classical optimization methods cannot be used. In this study Differential Evolution algorithms (DEs) \cite{storn_usage_1996} were used, which are inspired in the natural evolution. 

The main parts of DE algorithms are: (1) initialization, (2) evaluation, (3) mutation, (4) crossover, and (5) selection. Proposed algorithm's pseudocode is depicted in Algorithm~\ref{alg:opt-general}.

To adapt the DE algorithm to a constrained optimization problem, those constraints were incorporated into the DE framework through a constraint-handling mechanism.  Besides, a carefully designed initialization technique was employed to generate an initial population which taked into account that the problem formalization was done using historical data. In the following, adopted strategies for DE in this work are explained.\\

\begin{algorithm}[H]
\caption{Data-Driven Differential Evolution with Multi-Level Penalty Functions and Surrogate Models}
\label{alg:opt-general}
\begin{algorithmic}[1]
\State \textbf{Input:} Data path, product type, population size $n_{\text{pop}}$, max iterations $max\_iter$, mutation rate $F$, crossover rate $Cr$, mutation type, $I_c$, relative tolerance $\delta$
\State \textbf{Initialize} population $P \gets$ \texttt{get\_n\_samples}(data\_path, product\_type, $n_{\text{pop}}$, $I_c$)
\State Evaluate initial fitness $f_P \gets \texttt{fitness}(P, product\_type,  n_{\text{pop}}, \delta)$
\State Store initial solutions and compute dispersion \texttt{dispersion}$(P)$

\For{$iter = 1$ to $max\_iter$}
    \State $V_G \gets$ \texttt{mutation}($P$, $F$, mutation\_type)
    \State $U_G \gets$ \texttt{crossover}($P$, $V_G$, $Cr$)
    \State Adjust $U_G$ to variable bounds using \texttt{check\_ranges(data\_path, product\_type)}
    \State Evaluate offspring fitness $f_U \gets \texttt{fitness}(U_G, product\_type,  n_{\text{pop}}, \delta)$

    \For{each individual $i = 1$ to $n_{\text{pop}}$}
        \If{$f_{U_i} < f_{P_i}$}
            \State Replace $P_i \gets U_i$
            \State Update fitness $f_{P_i} \gets f_{U_i}$
        \EndIf
    \EndFor

    \State Update best solution and compute new dispersion \texttt{dispersion}$(P)$
    \State Save current population and fitness values for iteration $iter$
\EndFor

\State \textbf{Output:} Final fitness values $f_P$, final population $P$, full solution history, full dispersion history
\end{algorithmic}
\end{algorithm}

\subsubsection{Data-Driven Population Initialization}

Most population-based metaheuristics are initialized using a uniform distribution for the initial population creation \cite{ahmad_differential_2022}. However, this initialization considers that all individuals of the search space are equally probable, which could conduce the optimization to irrelevant zones and slow down the optimization \cite{poikolainen_cluster-based_2015}.

Instead, this work combined two techniques to create the initial population. The ratio of individuals created with each technique was governed by the \textit{Ic} parameter. 

\begin{itemize}
    \item \textbf{Best historical individuals (1 - Ic)}:
    Consisting of starting the search from the best individual recorded in the extrusion database. Which had the risk of prematurely converging in a sub-optimal solution. 
    \item \textbf{Parameter distribution fitted individuals (Ic)}: A joint distribution of the decision space was fitted to the data to maintain a balance between convergence and diversity in the population. A Gaussian Mixture model was fitted to the decision space of each product type. The data points that correspond to the same Gaussian distribution and the parameters for each Gaussian were fitted by the Expected-Maximization algorithm (EM) \cite{dempster_maximum_1977}. Data were scaled before the joint distribution approximation using a standard scaler and Scikit-learn implementations were used for all these tasks.
\end{itemize}
The combination or hybridation of both techniques is illustrated in part (1) of Figure~\ref{fig:MetaheuristicOverview}. For a more detailed explanation of the population initialization the pseudocode is described in Algorithm~\ref{alg:getnSamples}.

\begin{algorithm}[H]
\caption{Initial Population Sampling (\texttt{get\_n\_samples})}
\label{alg:getnSamples}
\begin{algorithmic}[1]
\State \textbf{Input:} Data path, product type, population size $n_{\text{pop}}$, density rate $I_c$
\State $n_{\text{density}} \gets \texttt{round}(n_{\text{pop}} \cdot I_c)$
\State $n_{\text{best}} \gets n_{\text{pop}} - n_{\text{density}}$
\State Read historycal data data $\gets$ \texttt{ReadData}(data\_path)
\State Filter by product type filtered\_data $\gets$ \texttt{FilterByProduct}(data, product\_type)
\State Scale data (scaled\_data, scaling\_params) $\gets$ \texttt{ScaleData}(filtered\_data)
\State Join distribution estimation kernel\_estimation $\gets$ \texttt{EstimateGaussianKernel}(scaled\_data)
\State Get density population density\_pop $\gets$ \texttt{SampleFromKernel}(kernel\_estimation, $n_{\text{density}}$)
\State Get best historycal population best\_pop $\gets$ \texttt{SelectBestHistoricalPoints}(filtered\_data, $n_{\text{best}}$)
\State Unscale population unscaled\_density\_pop $\gets$ \texttt{UnscaleData}(density\_pop, scaling\_params)
\State Merge population initial\_population $\gets$ \texttt{MergeSamples}(best\_points, unscaled\_density\_pop)
\State \textbf{Output:} initial\_population
\end{algorithmic}
\end{algorithm}

\subsubsection{Mutation, crossover, and selection}

According to \cite{deng_differential_2021} there are three main common mutation strategies:
\begin{itemize}
\item Random strategies (\textit{Rand}): Where random individuals are used to compute the mutation.
\item Best individual strategies (\textit{Best}): Where both random individuals and the current best individual are used for mutation.
\item Current to random strategies (\textit{Current-to-rand}): Where each individual with the combination of other random individuals are used to compute the mutation.
\end{itemize}
 
The mutation of DE algorithms is computed by a weighted sum of each individual and the difference between those selected by the corresponding strategy. In all the cases, a uniform distribution was used to determine the random individuals. In addition, the number of individuals used for mutation can vary (represented by the number following the mutation strategy).
 
From the set of most used strategy combinations suggested by \cite{deng_differential_2021} (DE/rand/1, DE/rand/2, DE/best/1, DE/best/2, and DE/current-to-rand/1) it was decided to discard DE/best/1 and DE/best/2 following the findings of \cite{sun_adaptive_2020}, given their trend to reach premature convergence into sub-optimal solutions.
Considering the previous, this study compared the first variants of rand and current-to-rand strategies.

In the case of crossover and selection methods, original algorithm methods with static crossover and selection parameters were used. These parts of the optimization algorithm are explained in the $3$rd and the $6$th scheme numbers of Figure~\ref{fig:MetaheuristicOverview}.

\subsubsection{Multi-Level Penalty Functions with Surrogate Constraints}

By design, DE algorithms are not able to handle constrained problems. In this work penalty functions were used to cope with the constraints.

Considering an equal penalty for all infeasible samples regardless of their degree of severity may lead to convergence towards sub-optimal solutions and fail to maintain a balance between exploration and exploitation of the algorithm. Therefore, this paper proposes a penalty function that uses the prediction of the surrogate constraint model to define a multi-level function.

As explained before, we have two types of constraints. For the model that mimics the visual inspection, the predicted class was used for the definition of the penalty function. Let $\vb{x}$ be an individual of the population in a fixed generation, and $g(\vb{x})$ be the predicted class of the visual inspection model. $g(\vb{x})$ takes values from $\{1, 2, 3\}$ that correspond to the steady, non-steady, and rejected classes, respectively. As the only feasible solutions are the $\vb{x}$ individuals fulfilling $g(\vb{x}) = 1$, the multi-level penalty function was defined as follow,

\begin{equation}
    \hat{y}'(\vb{x}) = 5(g(\vb{x}) + 1)^{g(\vb{x}) + 1}\hat{y}(\vb{x}),
\end{equation}
where $\hat{y}(\vb{x})$ is the predicted objective value of $\vb{x}$ and $\hat{y}'(\vb{x})$ is the correction done by the penalty function to $\hat{y}(\vb{x})$ when $g(\vb{x}) = 2, 3$. The penalty function was defined to ensure that for every $\vb{x}_1$ fulfilling $g(\vb{x}_1) \neq 1$, $\hat{y}'(\vb{x}_1) > \hat{y}(\vb{x}_2)$ for all $\vb{x}_2$ individuals with $g(\vb{x}_2) = 1$.

In the case of the solution consistency constraint, $\vb{x}$ individuals not fulfilling Equation~\ref{eq:phy-ctr} (using the previously penalized $\hat{y}'$), the predicted objective value $\hat{y}'$ was corrected as follows,

\begin{equation}
    \hat{y}'' = t_{setpoint} + 10,
\end{equation}
where $t_{setpoint}$ is the time required to have the extruders speeds in the defined setpoint which is defined in Equation~\ref{eq:t-set}.

Finally, for those $\vb{x}$ individuals that fulfill all the constraints of the problem, the objective value was defined by the XGBoost model's prediction, that is, $\hat{y}$.\\

The resulting penalized fitness value is depicted in the $5$th part of the Figure~\ref{fig:MetaheuristicOverview}. For a more detailed description, Algorithm~\ref{alg:multiLevelPenalty} shows the pseudocode for the multi-level penalty function proposed in this work.

\begin{algorithm}[H]
\caption{Multi-Level Penalty with Surrogate Constraints}
\label{alg:multiLevelPenalty}
\begin{algorithmic}[1]
\State \textbf{Input:} Population $P$, population size $n_{\text{pop}}$, constraint surrogate model $M_c$, approximate objective values $\hat{y}_P$, relative tolerance $\delta$
\For{every individual $i = 1$ to $n_{\text{pop}}$}
    \State $P_i \gets$ individual $i$ in population $P$
    \State Evaluate minimum quality constraint $g_{P_i} \gets$ \texttt{PredictQualityConstraint}($P_i$, $M_c$)
    \State Add Multi-level penalty to infeasible solutions $\hat{y}'_{P_i} \gets 5 \cdot (g_{P_i} + 1)^{g_{P_i} + 1} \cdot \hat{y}_{P_i}$
    \State Calculate time requiered to setup $t_{\text{setpoint}} \gets \max \left\{ P_{i,\text{speed},k}/P_{i,\text{accel},k} \mid k = 1,\dots,5 \right\}$ 
    \If{$t_{\text{setpoint}} < (1 + \delta)\cdot \hat{y}'_{P_i}$}
        \State $\hat{y}''_{P_i} \gets \hat{y}'_{P_i}$
    \Else
        \State Add penalty for inconsistent solution $\hat{y}''_{P_i} \gets t_{\text{setpoint}} + 10$
    \EndIf
    \State Store penalized objective $\hat{y}''_{P_i}$ for individual $i$
\EndFor
\State \textbf{Output:} Penalized objective values $\hat{y}''_P$
\end{algorithmic}
\end{algorithm}

Once the multi-level penalty mechanism has been established, it is integrated into the overall fitness evaluation process.
For that, the objective function's surrogate model is used to approximate the objective value of each individual. Then, first surrogate's output is corrected using the penalty function to account for constraint violations and ensure solution consistency. The complete procedure for estimating the penalized fitness values is presented in Algorithm~\ref{alg:fitness}, which is also described visualy in the steps from 3 to 5 of Figure~\ref{fig:MetaheuristicOverview}.

\begin{algorithm}[H]
\caption{Fitness Estimation with Surrogate-Based Objective estimation and Multi-Level Penalty}
\label{alg:fitness}
\begin{algorithmic}[1]
\State \textbf{Input:} Population $P$, product type, population size $n_{\text{pop}}$, relative tolerance $\delta$
\State Load surrogate model $\hat{f}$ trained for objective approximation
\State Load surrogate model $M_c$ trained for minimum quality constraint prediction
\State Predict approximate objective values $\hat{y}_P \gets \hat{f}(P, \text{product\_type})$
\State Apply multi-level penalty $\hat{y}''_P \gets$ \texttt{MultiLevelPenalty}($P$, $n_{\text{pop}}$, $M_c$, $\hat{y}_P$, $\delta$)
\State \textbf{Output:} Penalized fitness values $\hat{y}''_P$
\end{algorithmic}
\end{algorithm}

\section{Results}\label{sec:results}

\subsection{Constraints Identification Results}

A random partition was used to provide an unbiased estimation of the model's performance and prevent overfitting issues, allocating 80\% of the data for training and the remaining 20\% for testing. The model was validated on the training set by a 5-fold stratified cross-validation strategy.

Besides, a hyperparameter selection was performed by a grid search on the validation. The considered hyperparameters were the number of minimum samples per leaf (5 values tested from 5 to 100), and the minimum samples needed to split a node (all values tested from 2 to 7). The optimal parameters were finally set to 5 and 7, respectively. 

In the training phase, the F1-score's mean was 0.84, indicating a high overall performance of the model. Additionally, the standard deviation in the F1 metric of 0.04 suggests robustness, as it indicates minimal variation in scores around the mean. For the testing, the F1-score value was 0.87.

\begin{figure}
    \begin{center}
           \includegraphics[scale = 0.35]{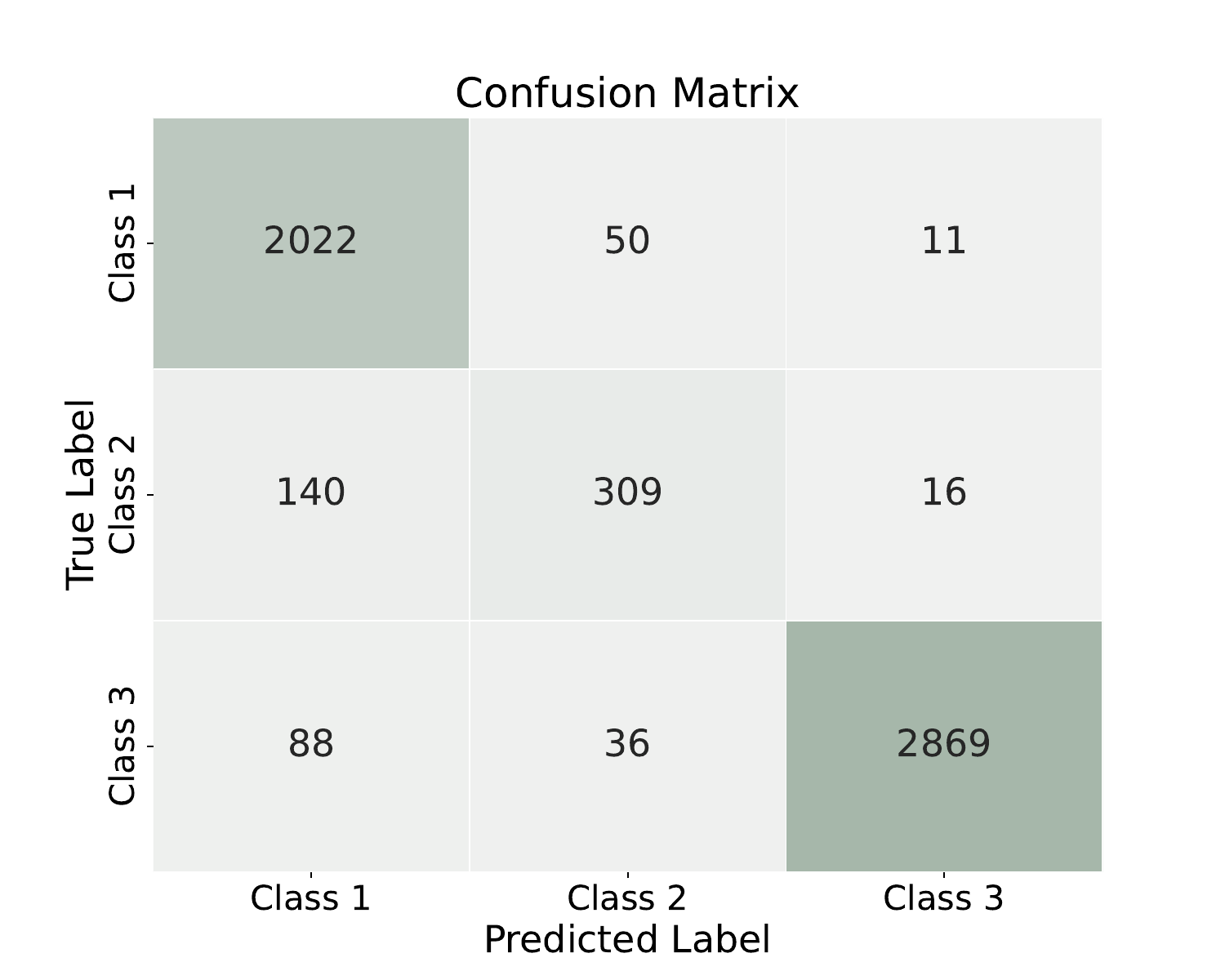} 
           \caption{Confusion matrix for the constraint surrogate model in the final model}
           \label{fig:cm}
    \end{center}
\end{figure}

As shown in Figure~\ref{fig:cm}, the class with the worst classification performance on the model was the not steady class, which was expected because of the heterogeneity in this
class (not all were equally far from the defined interval near the target profilometer value).

\subsection{Results of the Surrogate Objective Function}
From the whole dataset, the $80\%$ of random samples were used for the training, and the remaining $20\%$ of data were been used for testing. The validation was done on the training set by a 5-fold cross-validation. 

For the hyperparameter selection, some hyperparameters were tuned. Three of them were selected to control the growth of the trees; 

\begin{itemize}
    \item Maximum depth of the trees: 4 values tested (ranging from $2$ to $5$).
    \item Minimum loss reduction for further splits in a node or \textit{gamma} parameter: 7 values tested (ranging from $0$ to $8$).
    \item Number of tree estimators or boosted trees: 7 values tested (ranging from $5$ to $150$).
\end{itemize}

The remaining ones were selected to control the boosting.

\begin{itemize}
    \item Features weight shrinkage or \textit{eta} parameter: 5 values tested (ranging from $0.1$ to $0.7$).
    \item Sampling method: tested with uniform and gradient-based methods.
    \item Sampling ratio to be used in the boosting: 5 values tested (ranging from $0.1$ to $1$).
\end{itemize}

Finally, found the best hyperparameters were $3$, $0.7$, $50$, $0.1$, uniform sampling and $1$, respectively. The mean Kendall correlation value obtained in the validation for those hyperparameters was 0.41 and the standard deviation 0.03. 

\begin{figure}
    \begin{center}
        \centering
           \includegraphics[scale = 0.4]{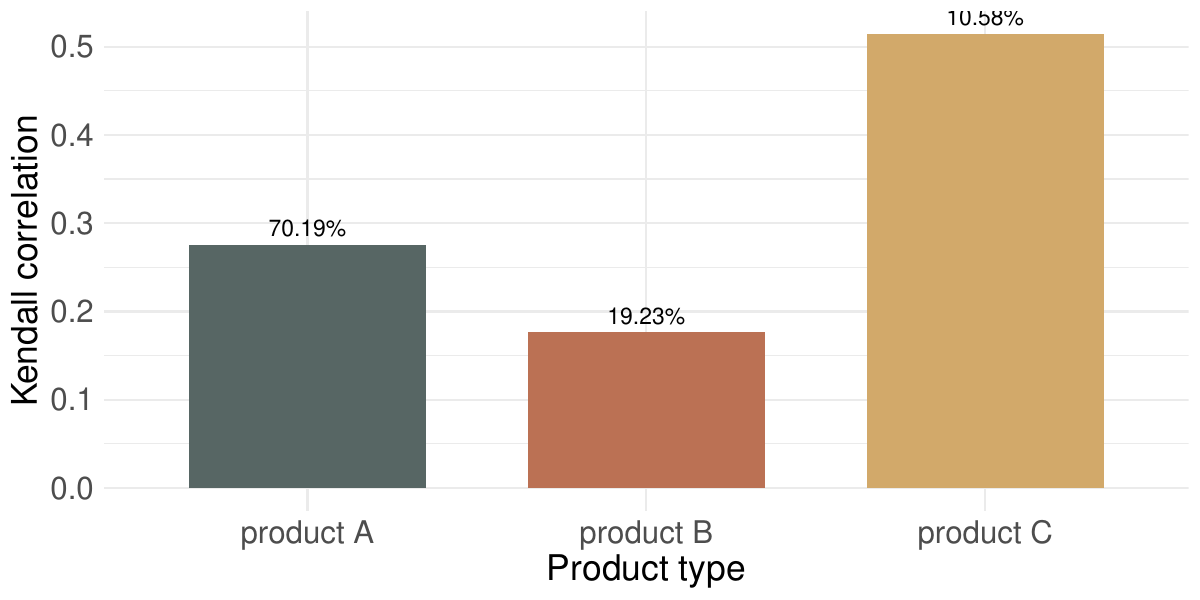}
           \caption{Kendall Correlation results and usage percentage for the test and product type. }
           \label{fig:M2_results}
    \end{center}
\end{figure}

The overall Kendall correlation value in the testing was 0.41 but values differ for the different product types. Figure~\ref{fig:M2_results} shows the obtained values in the testing for each product type, where the percentage of use of each product is also illustrated (69.7\%, 19.6\% and 10.7\% respectively for A, B and C). As can be seen, the C product type's correlation is especially higher than A and B, even with a lower use percentage.

\begin{figure}
    \begin{center}
           \includegraphics[scale = 0.5]{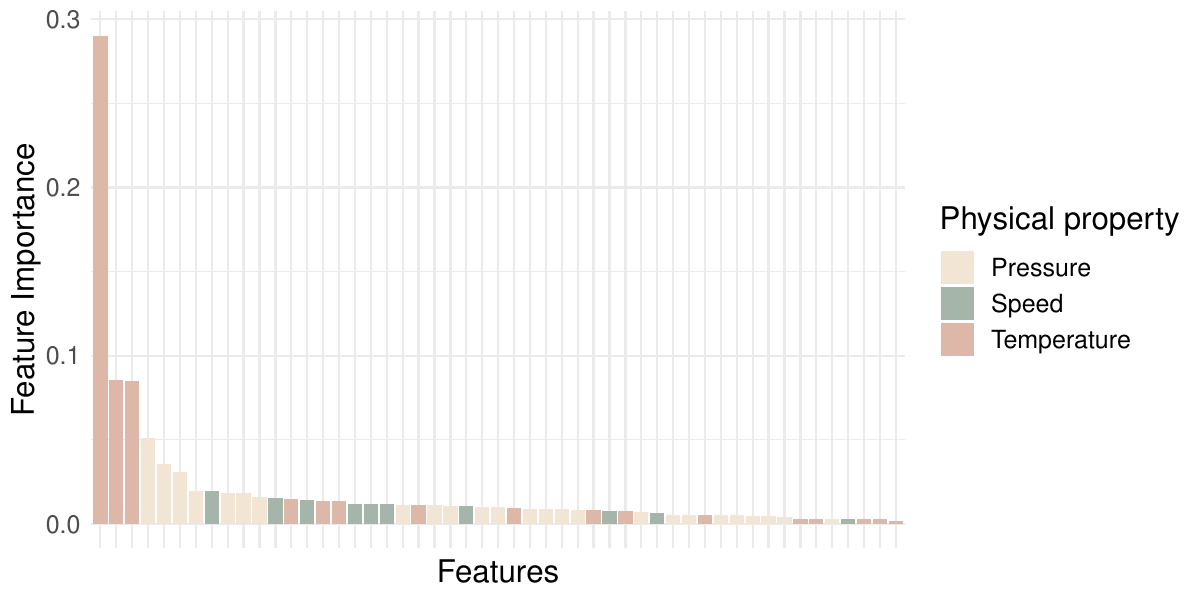}
           \caption{Variable importances by measured physical properties of the XGBoost model for objective function approximation.}
           \label{fig:M2_importances}
    \end{center}
\end{figure}

To know which variables were the most influential in the model, XGBoost's model importance was used. Figure~\ref{fig:M2_importances} shows the importances of the variables categorized by the physical property that are measuring. 

The seven most influential variables represented a sum of 0.6 decision importance, while the remaining 0.4 importance was spread through more than 40 variables. That means that the decision variables were interacting with each other, and the combination of many variables was needed to predict the estimated \textit{steadiness time} in the profilometer. Furthermore, it can be seen that the most influential variables were related to the temperature and that, in general, the only influential parameters were related to the physical properties, that is, the product type used was not meaningful in the decision.

\subsection{Optimization results}
The selection of the parameters for Differential Evolution algorithms is crucial for the convergence of them. 

Is been shown that a too small population can lead to bad results because of the low diversity while having a too large population, the algorithm can be inefficient \cite{ang_constrained_2020}. In our work, the solution presented in \cite{yu_data-driven_2022} was followed, and the population size was set to $100$ individuals.

Following \cite{jaber_branch-and-bound_2022} and \cite{mai_machine_2021}, values of the \textit{mutation parameter} and \textit{crossover rate} were set to be higher to $0.5$. For the first, $\{0.7, 0.8, 0.9\}$ values were used, and for the second, $\{0.5, 0.7, 0.9\}$ values were tested.

For the maximum number of iterations, the complexity of the problem as well as the computational cost was taken into account. The maximum number of iterations was set to $500$, for each of the combinations of the parameters and mutation strategy.

Finally, the combination of initialization was set to $0.25$, that is the 25\% of the individuals were initialized by the estimated distribution, whereas, the remaining 75\% was initialized by the corresponding best historical results.

To summarize the experimentation done, Table~\ref{table:tabla_ids} defines the used identification number for each combination of parameters in our DE algorithm. The identification is identical for all maximum numbers of iterations and product types. To assess the stability of the results, each of the experiments was repeated $10$ times.

\begin{table}
\centering
\begin{tabular}{ | m{0.5cm} | m{3.9cm}| m{0.5cm} | m{0.5cm} || m{0.5cm} | m{3cm}| m{0.5cm} | m{0.5cm} |} 
\hline
\textbf{ID} & \textbf{Mutation} & \textbf{F} & \textbf{Cr} & \textbf{ID} & \textbf{Mutation} & \textbf{F} & \textbf{Cr}\\
\hline
\textbf{1} & current-to-rand/1 & 0.7 & 0.5 & \textbf{10} & rand/1 & 0.7 & 0.5\\
\hline
\textbf{2} & current-to-rand/1 & 0.8 & 0.5 & \textbf{11} & rand/1 & 0.8 & 0.5\\
\hline
\textbf{3} & current-to-rand/1 & 0.9 & 0.5 & \textbf{12} & rand/1 & 0.9 & 0.5\\
\hline
\textbf{4} & current-to-rand/1 & 0.7 & 0.7 & \textbf{13} & rand/1 & 0.7 & 0.7\\
\hline
\textbf{5} & current-to-rand/1 & 0.8 & 0.7 & \textbf{14} & rand/1 & 0.8 & 0.7\\
\hline
\textbf{6} & current-to-rand/1 & 0.9 & 0.7 & \textbf{15} & rand/1 & 0.9 & 0.7\\
\hline
\textbf{7} & current-to-rand/1 & 0.7 & 0.9 & \textbf{16} & rand/1 & 0.7 & 0.9\\
\hline
\textbf{8} & current-to-rand/1 & 0.8 & 0.9 & \textbf{17} & rand/1 & 0.8 & 0.9\\
\hline
\textbf{9} & current-to-rand/1 & 0.9 & 0.9 & \textbf{18} & rand/1 & 0.9 & 0.9\\
\hline
\end{tabular}
\caption{\label{table:tabla_ids} Identification for each combination of mutation strategy, mutation operator, and crossover rate. Applies to all experiments with different product types.}
\end{table}

\begin{figure}[t]
    \begin{center}
           \includegraphics[scale = 0.45]{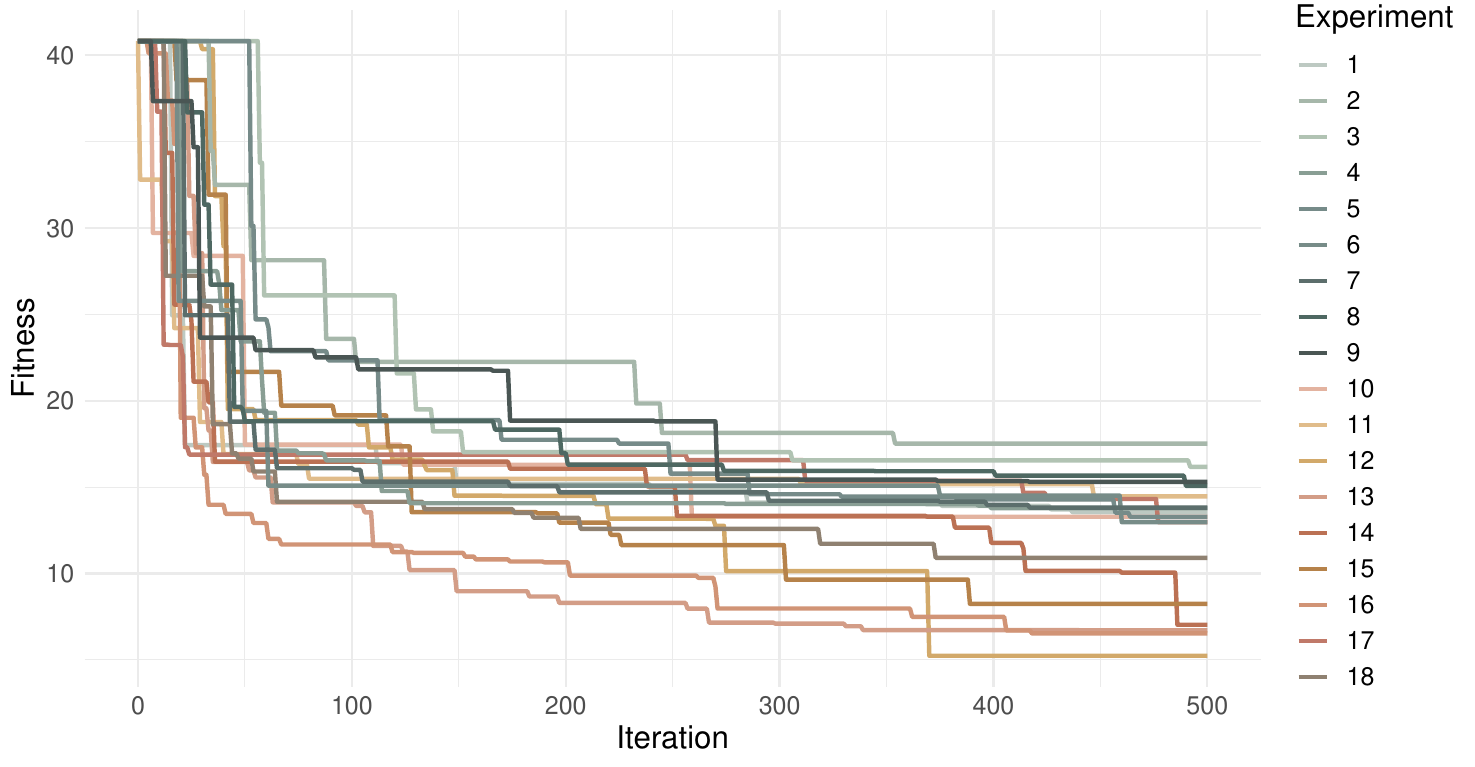}
           \caption{Fitness convergence curves for product type A and maximum number of iterations $500$.}
           \label{fig:Convergence_0100000099}
    \end{center}
\end{figure}

Analyzing the convergence of the algorithms, it has been seen that with the experimentation done, $150$ iterations are not enough to have a convergence in the optimal solution search. This is shown in Figure~\ref{fig:Convergence_0100000099} for most used product type, A. However, all the experiments converge with $300$ iterations, and no significant changes have been seen with $500$ iterations. Besides, fixing the maximum number of iterations in $300$ and not in $500$ could save $40\%$ of the computation cost. Therefore, we conclude that the best results are obtained by setting the maximum number of iterations to $300$.

\begin{figure}[t]
    \begin{center}
           \includegraphics[scale = 0.55]{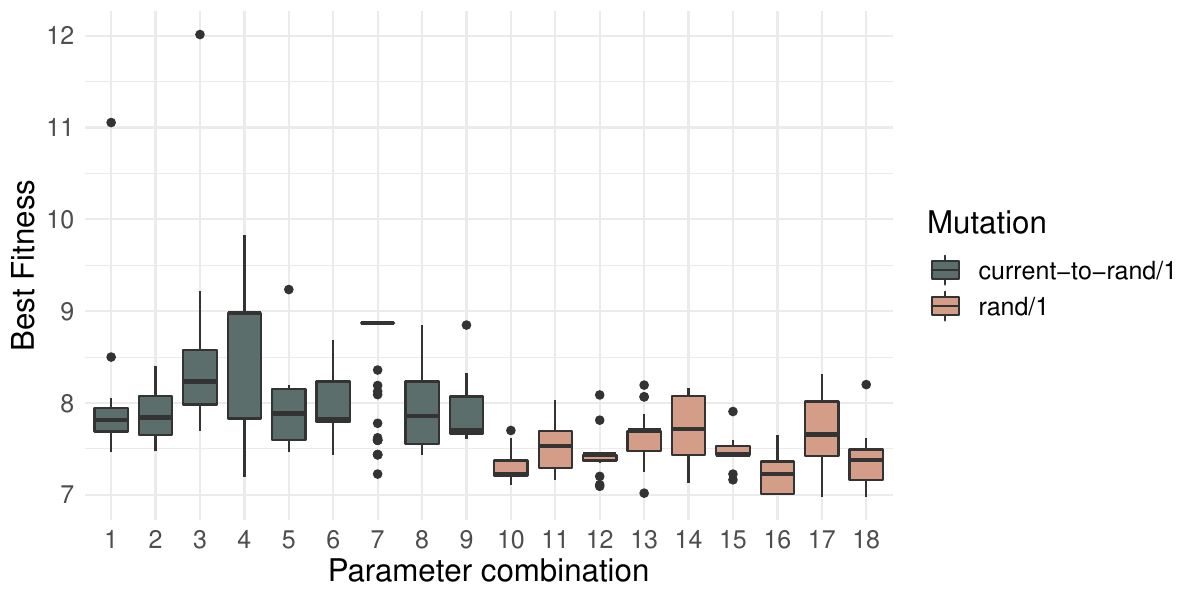}
           \caption{Boxplots for the results with product type $C$ and 300 iterations, after 10 algorithm runs.}
           \label{fig:boxplots_0100000072}
    \end{center}
\end{figure}

Within the used mutation strategies, is been seen that the rand/1 strategy gets better results than the current-to-rand/1. In our case, the initialization is made by the combination of the 75\% of best historical results and 25\% created by the fitted distribution. With such initialization, rand/1 is able to perform a more extensive exploration in the decision space than current-to-rand/1 strategy, which probably leads to the search getting stuck in a sub-optimal region. Figure~\ref{fig:boxplots_0100000072} shows that best results with $300$ iterations and product type $C$ after 10 runs of the algorithm got better results for experiments with rand/1 strategy than with current-to-rand/1 strategy.

\begin{figure}
    \begin{center}
           \includegraphics[scale = 0.45]{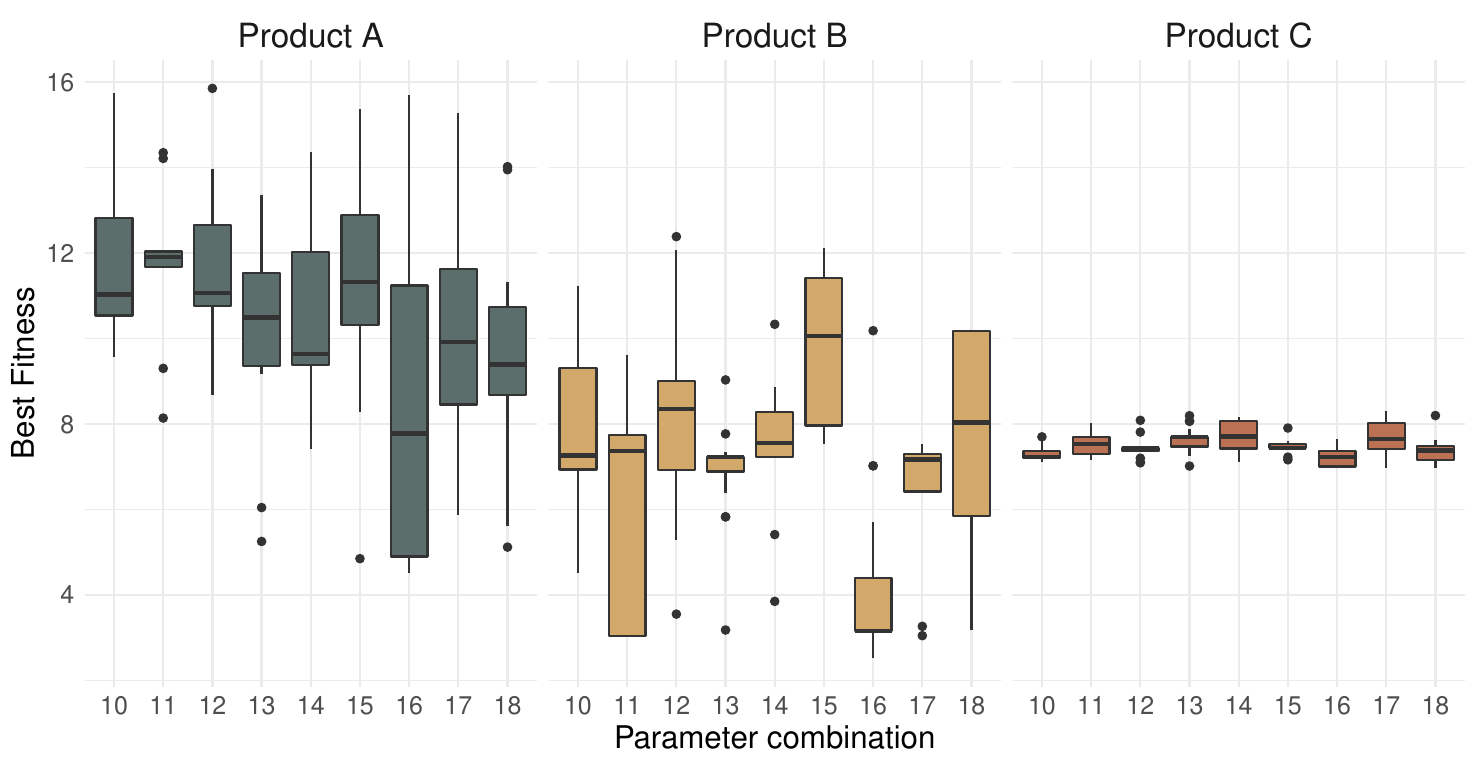}
           \caption{Boxplots for the results with 300 iterations and rand/1 mutation strategies for the three product types.}
           \label{fig:boxplots_all}
    \end{center}
\end{figure}

For all the products, the global minimum has been achieved by the $16$th combination, that is by setting $F = 0.7$ and $Cr = 0.9$, but the variability in all the attempts is high in some cases. The variability within the 10 runs reflects, in part, the uncertainty of the model that approximates the objective function, but also depends on the global search capacity of the optimization algorithm. The variability observed in Figure~\ref{fig:boxplots_all}, together with the results obtained when fitting the objective function by product type (see Figure~\ref{fig:M2_results}), shows that the model’s uncertainty contributes to the variability in the optimization outcomes, with more stable results achieved for product type $C$. For the most used product, i.e. product $A$, the variability of the $16$th combination is huge compared with other experiments like the $17$th or $18$th, which have achieved similar results if all attempts are taken into account. In the case of product $B$, the variability of the $16$th experiment is low, thus, unquestionably, it is the best parameter combination in this case. Regarding to the last product type, product $C$, the $16$th combination has achieved the global minimum, but combination $10$ achieves similar results with less variability in all attempts. 

To ensure that good results are obtained regardless of the number of trials conducted, it is better to select parameter combinations that achieve low variability, even if the global minimum is achieved with another combination that has higher variability.

\section{Discussion}\label{sec:discussion}
\begin{figure}
    \begin{center}
           \includegraphics[scale = 0.45]{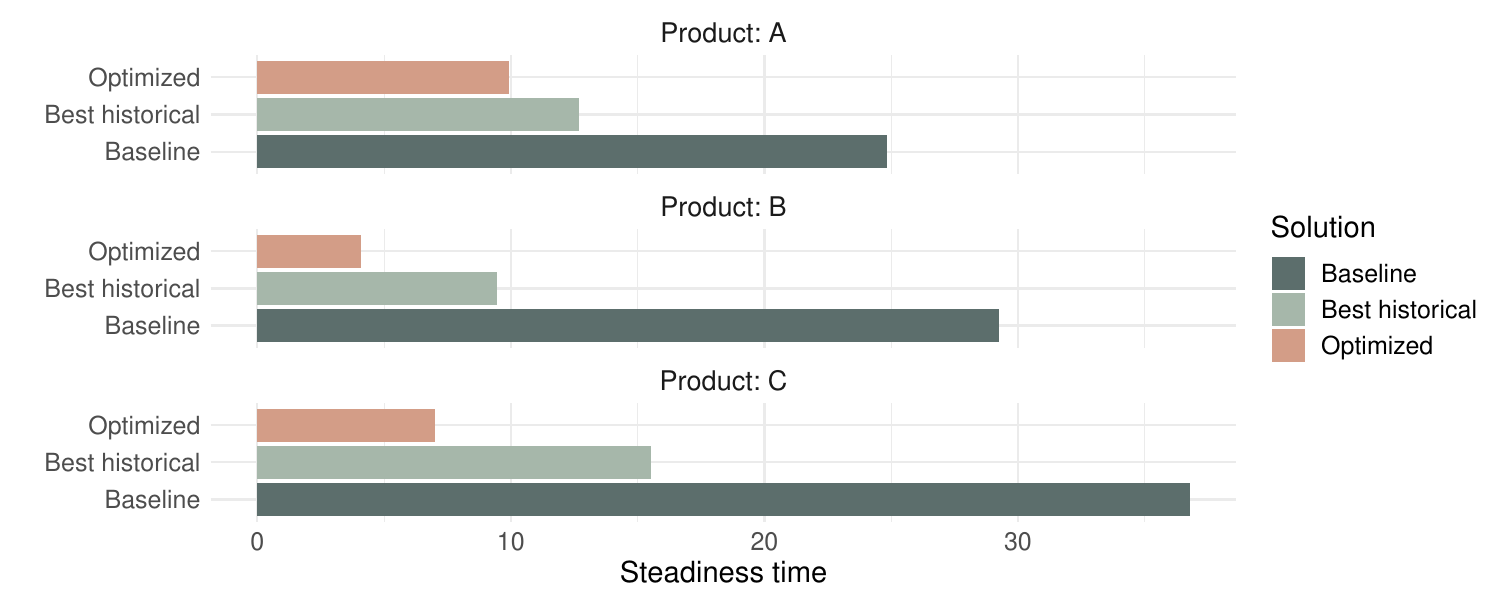}
           \caption{Baseline reference, best historical values after mathematical formulation and optimal results of the best solutions for each of the products.}
           \label{fig:discussion}
    \end{center}
\end{figure}

In order to assess the improvement of the presented proposal to achieve the problem objectively, both the improvements made by the mathematical formulation and by further optimization are compared to the initial baseline of the problem. These improvements are visualized in Figure~\ref{fig:discussion}.

The problem studied lacked a mathematical formulation as an optimization problem, nor was there a measure of quality for each of the extrusions or tests carried out in the process, we have quantified the goodness of the baseline solutions in the fairest way.

As there is no reference configuration for each of the product types used, as these depend on each case and on the person in charge, an approximation of the average configuration is made. For that, the medians of the speed setpoints were calculated for each extruder and product type, for those extrusions that got steady in the profilometer. Finally, the time to steadiness was predicted by the trained XGBoost model, so the initial baseline was calculated. The obtained baseline results were $24.82$ seconds, $29.25$ seconds, and $36.78$ seconds for products A, B and C, respectively. These values are represented in Figure~\ref{fig:discussion} with the \textit{baseline} bar for the three products.

As previously mentioned, the mathematical formulation of the optimization problem using the provided surrogate models results in a historical dataset with suboptimal solutions. The best solutions found for each product type are used as starting points in the optimization algorithm, as mentioned in Section~\ref{sec:DE}. Therefore, the mathematical formalization leads to an improvable suboptimal solution of $12.70$s for product A, $9.47$s for product B, and $15.52$s for product C, as it is visualized in Figure~\ref{fig:discussion} with \textit{best historical} bars. At this point, it is clear that the mere mathematical formulation of the problem is capable of improving the process baseline.

When comparing these suboptimal solutions to the optimized results presented in the results section, the advantages of the approach become more apparent. The optimized times achieved are $9.93$s for product A, $4.10$s for product B, and $7.00$s for product C. These values are illustrated in the Figure~\ref{fig:discussion} with \textit{optimized} bars. This notable improvement underscores the potential of the presented surrogate based optimization algorithm. The optimized results not only surpass the baseline but also show a considerable enhancement over the initial suboptimal solutions, suggesting the strength and applicability of the proposed optimization framework.

Finally, if the presented solution is used in the real process it is expected that the mean reduction regarding the \textit{steadiness time} is about 60\% in the case of product A, 86\% in the case of product B and 81\% in the case of product C. If the product distribution is maintained, this improvement would have an average total reduction of 65\% in the time needed to get the product's quality steady. Moreover, the solutions obtained for each of the products could also accelerate the setup process, so the time required to start up the extruders can also be reduced. 

\section{Conclusions}\label{sec:conclusions}

The Surrogate-model based optimization formulation and the Data-Driven Differential Evolution with Multi-Level Penalty Functions have collectively led to significant improvements in the process baseline for all product types studied. Initially, the mathematical formulation reduced the time to steadiness for each product type, while the subsequent optimization algorithm further enhanced these results. This efficient optimization approach demonstrates the robustness and effectiveness of the proposed framework in addressing the tire industry process optimization, resulting in substantial process improvements and increased operational efficiency.

By integrating data-driven models and metaheuristic algorithms within the surrogate optimization paradigm, the paper offers a practical approach to modeling and optimizing complex industrial problems with the only use of historical data. This underlines its relevance and applicability in real-world manufacturing environments, as well as opening up the research avenue proposed in the literature.

However, it is noteworthy to mention that, even if the optimization presented does use Machine Learning models for process surrogate, the intrinsic prediction errors in which these models incur have been not fully contemplated. This uncertainty can play an important role when working with historical data only, as it is not possible to work with exploration and exploitation methods that can minimize model error. This aspect is known in the literature as quantification of uncertainty or robust optimization \cite{parnianifard_robust_2021}, \cite{wang_surrogate-assisted_2020} and could be of interest to consider in the future to improve the current approach.

The authors acknowledge the vast research landscape of metaheuristic algorithms, recognizing that alternative approaches may yield even better results than those presented here. Nonetheless, it is emphasized that the objective of this work has not been to exhaustively explore such alternatives but rather to stimulate further exploration and comparison of different approaches in this regard. Moving forward, it would be of interest to continue exploring and comparing different alternatives in the context of metaheuristic algorithms. 

Finally, so that the work can be exploited in the real industrial process, and thus be possible to validate and optimize it with the help of the operators, it would be interesting to treat the optimization and the proposed models as a Digital Twin. In that case, the proposed approach could be improved with continuous learning of the process.

\vspace{6pt} 

%%%%%%%%%%%%%%%%%%%%%%%%%%%%%%%%%%%%%%%%%%
%% optional
%\supplementary{The following supporting information can be downloaded at:  \linksupplementary{s1}, Figure S1: title; Table S1: title; Video S1: title.}


% Only used for preprtints:
% \supplementary{The following supporting information can be downloaded at the website of this paper posted on \href{https://www.preprints.org/}{Preprints.org}.}

%\begin{tabularx}{\textwidth}{lll}
%\toprule
%\textbf{Name} & \textbf{Type} & \textbf{Description} \\
%\midrule
%S1 & Python script (.py) & Script of python source code used in XX \\
%S2 & Text (.txt) & Script of modelling code used to make Figure X \\
%S3 & Text (.txt) & Raw data from experiment X \\
%S4 & Video (.mp4) & Video demonstrating the hardware in use \\
%\bottomrule
%\end{tabularx}
%}

%%%%%%%%%%%%%%%%%%%%%%%%%%%%%%%%%%%%%%%%%% 
\section*{Author Contributions}
Eider Garate-Perez and Kerman López de Calle-Etxabe have contributed to conceptualization, methodology, investigation and to the writing of the original draft and review. Eider Garate-Perez has also contributed to the software, validation, formal analysis, data curation and visualization. Finally, Susana Ferreiro has contributed to the conceptualization, and to the review of the writing.

\section*{Funding}
This work is been done under AI-PROFICIENT project which has received funding from the European Union's Horizon 2020 research and innovation program under grant agreement No. 957391.
This work was further supported by the Basque Government BEREZ-IA ELKARTEK 2023 Programme under grant agreement No KK-2023/00012].

\section*{Data Availability}
Continental France SAS has provided domain knowledge and historical data for this study. The datasets generated and/or analyzed during the current study are not publicly available due to confidentiality agreements and commercial restrictions. The data belong to Continental SAS France and contain proprietary information.

\section*{Acknowledgments}
The authors would like to thank Dr. Borja Calvo for his valuable contributions and insightful discussions during the course of this work.

\section*{Conflicts of Interest}
The authors declare no conflicts of interest.

\bibliographystyle{ieeetr} 
\bibliography{references}

\end{document}